# Complex structure of turbulence across the ASDEX Upgrade pedestal

L.A. Leppin[1]†, T. Görler[1], M. Cavedon[2], M.G. Dunne[1], E. Wolfrum[1], F. Jenko[1] and the ASDEX Upgrade Team[3]

[1]Max Planck Institute for Plasma Physics, Boltzmannstr. 2, 85748 Garching, Germany
[2]Dept. of Physics "G. Occhialini", University of Milano-Bicocca, Milan, Italy
[3]See author list of U. Stroth et al. 2022 Nucl. Fusion 62 042006

The theoretical investigation of relevant turbulent transport mechanisms in H-mode pedestals is a great scientific and numerical challenge. In this study, we address this challenge by global, nonlinear gyrokinetic simulations of a full pedestal up to the separatrix, supported by a detailed characterisation of gyrokinetic instabilities from just inside the pedestal top to pedestal centre and foot. We present ASDEX Upgrade pedestal simulations using an upgraded version of the gyrokinetic, Eulerian, delta-f code GENE (genecode.org) that enables stable global simulations at experimental plasma $\beta$ values. The turbulent transport is found to exhibit a multi-channel, multi-scale character throughout the pedestal with the dominant contribution transitioning from ion scale Trapped Electron Modes (TEMs)/Micro Tearing Modes (MTMs) at the pedestal top to electron scale Electron Temperature Gradient modes (ETG) in the steep gradient region. Consequently, the turbulent electron heat flux changes from ion to electron scales and the ion heat flux reduces to almost neoclassic values in the pedestal centre. $E \times B$ shear is found to strongly reduce heat flux levels in all channels (electron, ion, electrostatic, electromagnetic) and the interplay of magnetic shear and pressure gradient is found to locally stabilise ion scale instabilities.

**Key words:** ETG, Turbulence, Gyrokinetics, GENE code, Pedestal, H-Mode

## 1. Introduction

The High-confinement mode (H-mode) of plasma operation in a tokamak fusion experiment is characterised by the presence of a radially narrow region just inside the Last Closed Flux Surface (LCFS, also called separatrix) in which turbulent transport is strongly suppressed. In this edge transport barrier the profiles of electron and ion temperature as well as density are steep and their gradients large, forming a pedestal that lifts the rest of the profiles, ultimately improving fusion conditions in the core of the experiment or future reactor. For this reason, H-mode is the baseline scenario of ITER, the world's largest fusion experiment, currently under construction in Cadarache, France.

In this study, we investigate small-scale turbulence in a pedestal of a standard Type-I ELMy H-mode from ASDEX Upgrade (AUG), which has been experimentally well diagnosed and analysed (Cavedon *et al.* 2017; Viezzer *et al.* 2020). Standard H-modes are unstable against Edge Localized Modes (ELMs), which expel large amounts of heat and particles ($\sim 10\%$ of plasma energy (Cathey *et al.* 2020)) tens of times per second, each time crashing the pedestal and causing unacceptably high heat loads on the divertor in future reactors. Therefore, special H-modes (EDA (Stimmel *et al.* 2022; Greenwald *et al.* 1999; Gil *et al.* 2020), QH (Burrell *et al.* 2005), small ELM/ QCE (Harrer *et al.* 2022)) and techniques (RMPs (Wade *et al.* 2015; Suttrop *et al.* 2011, 2018), vertical kicks (de la Luna *et al.* 2016), pellets (Baylor *et al.* 2015)) for the suppression or mitigation of ELMs are an area of very active research. Nonetheless, well-developed standard H-mode pedestals remain useful as a prototype to study what turbulent transport mechanisms persist in this region of strong turbulence suppression.

† Email address for correspondence: leonhard.leppin@ipp.mpg.de

Turbulent transport in the pedestal is a key ingredient in multiple important aspects of edge physics: It influences the pedestal height and width as well as the dynamics of ELM cycles and its suppression is the basis for the formation of the edge transport barrier.

The simulation and quantitative prediction of turbulent pedestal transport, however, is challenging. The steep gradients in temperature and density, which characterise the pedestal centre, drive many of the classic core instabilities at once, causing the parallel presence and interaction of different instability types (e.g. Ion Temperature Gradient (ITG) modes, Electron Temperature Gradient (ETG) modes and Trapped Electron Modes (TEMs)). Furthermore, the strong localisation of the drive poses fundamental challenges for heat flux calculations of ion scale turbulence with usual local gyrokinetic simulations in which the driving gradients are assumed to be constant over the simulation domain: The strong drive causes large eddies, which require wide simulation domains to prevent the eddy from "biting its own tail" across the periodic boundary condition, which would yield unphysically large heat fluxes. Since the gradients are assumed to be constant in the local approach, the total drive of the system in the local simulation ends up much higher (strong drive in the whole simulation domain possibly hundreds of gyroradii wide) compared to the very localised steep gradient region ($\sim$ten gyroradii) and may prevent convergence at all. Additionally, the steep gradients make the inclusion of electromagnetic fluctuations important, not only in order to include transport caused by electromagnetic instabilities (Kinetic Ballooning Mode (KBM), Micro Tearing Mode (MTM)) but also because they can significantly influence the level of electrostatic heat flux.

In recent years the following picture of pedestal micro-instabilities and transport has emerged: MTM and ETG have been identified to be strong candidates for the most relevant instabilities in the steep gradient region of the pedestal (Kotschenreuther *et al.* 2019; Hatch *et al.* 2019, 2016; Hassan *et al.* 2022; Halfmoon *et al.* 2022; Parisi *et al.* 2020; Chapman-Oplopoiou *et al.* 2022). In ASDEX Upgrade also TEMs and KBMs have been found to be relevant (Hatch *et al.* 2015) and in JET-ILW (ITER-like Wall) ITG has been found to be present (Hatch *et al.* 2017; Predebon *et al.* 2023). Most of these studies are based on local/linear, local/nonlinear, global/linear, or reduced $\beta$ simulations. However, to study which turbulent mechanisms drive transport under real pedestal conditions, global/nonlinear simulations at experimental $\beta$ values are indispensable. Such simulations are presented in this work.

In this paper, we present a radially resolved analysis of heat transport from pedestal top to foot covering ion and electron scales and including electrostatic as well as electromagnetic fluctuations. At the centre of our study are global, nonlinear, electromagnetic gyrokinetic simulations of an AUG pedestal, which have been enabled by an upgrade of the GENE code. They are supported by a detailed local, linear analysis of instabilities and dedicated local, nonlinear simulations of heat flux due to electron scale fluctuations. We find that turbulent ion heat flux is present on the pedestal top and vanishes towards the steep gradient region. This is observed even without including the effect of $E \times B$ shear due to a radial electric field $E_r$ but is strongly enhanced by it. A combination of magnetic shear and pressure gradient is identified to locally stabilise modes and suppress heat flux. Our findings are in agreement with experimental observations that the ion heat flux reduces to neoclassic levels in the pedestal centre. Turbulent electron heat flux remains approximately constant across the pedestal but changes scale from dominantly ion scale TEMs at the pedestal top to small-scale electron temperature gradient driven turbulence in the steep gradient region.

The paper is structured as follows: Sec. 2 introduces the GENE code and its new upgrade for nonlinear, electromagnetic, global simulations, Sec. 3 explains the experimental scenario investigated, Sec. 4 shows the results of linear, local scans to characterise instabilities, Sec. 5 shows the results of global and local nonlinear simulations and finally in Sec. 6 conclusions are drawn.

## 2. Upgrade of the global electromagnetic GENE code

All simulations presented in this paper have been performed with the GENE code (genecode.org) (Jenko *et al.* 2000; Görler *et al.* 2011), which is a gyrokinetic, Eulerian, $\delta f$ Vlasov code that can perform a variety of simulation types, including: linear, nonlinear, electrostatic, electromagnetic, global, local, neoclassic.

It was found that a certain type, namely nonlinear, electromagnetic, global simulations tended to be prone to numerical instabilities that could be avoided by artificially reducing plasma $\beta$ in simulations e.g. (Hatch *et al.* 2019). In effect, until now, nonlinear, global simulations could in most cases only be run with reduced $\beta$ and not with experimental values. In this section the upgrade that overcomes this numerical instability is presented, following the proof-of-principle in (Crandall 2019) based on (Reynders 1993). The same electromagnetic model has recently also been implemented in GENE-3D (Wilms *et al.* 2021), the stellarator version of the GENE code. Related approaches are also being followed in gyrokinetic particle-in-cell (PIC) codes (Mishchenko *et al.* 2017).

GENE solves the Vlasov-Maxwell system of equations to calculate the time evolution of a gyrocenter distribution $F$ in 5D phase space. The Vlasov equation for one of the species reads:

$$\frac{\partial F}{\partial t} + \frac{\partial \boldsymbol{X}}{\partial t} \cdot \nabla F + \frac{\partial v_\|}{\partial t} \frac{\partial F}{\partial v_\|} = C, \qquad (2.1)$$

where $\partial \boldsymbol{X}/\partial t$ contains the parallel and perpendicular (gradient-$B$, generalised $E \times B$, curvature) drifts $\partial \boldsymbol{X}/\partial t = v_\| \hat{\boldsymbol{b}} + B/B_\|^*(\boldsymbol{v}_{\nabla B} + \boldsymbol{v}_\chi + \boldsymbol{v}_c)$, $\partial v_\|/\partial t$ is the parallel acceleration and $C$ symbolises a collision operator. The parallel acceleration depends on the time derivative of the fluctuating part of the magnetic vector potential $\bar{A}_{1\|}$ (the overbar denotes a gyroaverage):

$$\frac{\partial v_\|}{\partial t} = -\frac{1}{mv_\|} \frac{\partial \boldsymbol{X}}{\partial t} \cdot \left( \mu \nabla (B + \bar{B}_{1\|}) + q \nabla \bar{\phi}_1 \right) - \frac{q}{mc} \frac{\partial \bar{A}_{1\|}}{\partial t}, \qquad (2.2)$$

where c is the speed of light, $q$ the charge and $m$ the mass of a given particle species. In the unmodified GENE code the time derivative of the fluctuating part $F_1$ and $\bar{A}_{1\|}$ are combined and the distribution function which is evolved in time is $g_1 = F_1 - (q/mc)\bar{A}_{1\|}\partial F_0/\partial v_\|$. This approach exhibits numerical instabilities in global, nonlinear, electromagnetic simulations. The numerical instabilities can be solved by retaining $F_1$ as the main distribution function and solving for $\bar{A}_{1\|}$ with an additional field equation derived from Ampère's law $\nabla_\perp^2 A_{1\|} = (-4\pi/c)j$. By applying a time derivative to Ampère's law, using $E_\|^{\mathrm{ind}} = (-1/c)\partial A_{1\|}/\partial t$ and writing the time derivative of $F_1$ as

$\partial F_1/\partial t = (q/m)\bar{E}^{\text{ind}}_{\|}\partial F_0/\partial v_{\|} + R$ one finds:

$$\nabla^2_{\perp} E^{\text{ind}}_{\|} = \frac{4\pi}{c^2}\sum_i q_i \int d^3v v_{\|} \mathcal{G}^{\dagger}\left\{\frac{q_i}{m_i}\bar{E}^{\text{ind}}_{\|}\frac{\partial F_{0,i}}{\partial v_{\|}} + R_i\right\}, \quad (2.3)$$

where the current $j$ has been expressed as a velocity space integral over the gyrocenter distribution function $F_1$, $R$ denotes all remaining terms, and the sum goes over all species $i$. Collecting all terms containing $E^{\text{ind}}_{\|}$ on one side of the equation the final field equation becomes:

$$\left(\nabla^2_{\perp} + \frac{4\pi}{c^2}\sum_i \frac{q_i^2}{m_i}\int d^3v \mathcal{G}^{\dagger} v_{\|}\frac{\partial F_{0,i}}{\partial v_{\|}}\mathcal{G}\right) E^{\text{ind}}_{\|} = \frac{4\pi}{c^2}\sum_i q_i \int d^3v \mathcal{G}^{\dagger} v_{\|} R_i, \quad (2.4)$$

which is solved numerically. Next to the additional field equation for the plasma induction this approach requires an additional nonlinear term between the fields. In total these changes increase the computational time per time step by approximately 30%. The new electromagnetic model is furthermore compatible with the use of block-structured velocity grids (Jarema *et al.* 2017).

Before we discuss the results obtained with the presented upgrade in global, nonlinear simulations in Sec. 5, the following two sections introduce the pedestal under consideration and its linear instabilities.

## 3. Experimental scenario: H-mode pedestal of ASDEX Upgrade #31529

The particularly well diagnosed and studied shot AUG #31529 (Cavedon *et al.* 2017; Viezzer *et al.* 2020) serves as the basis for our investigation. AUG #31529 has NBI (Neutral Beam Injection) and ECRH (Electron Cyclotron Resonance Heating) heating, with a total heating power of $P_{tot} = 8.7$ MW, an on-axis B-field of -2.5 T and a plasma current of 0.8 MA. From this shot, we employ ELM-synchronised profiles from (Cavedon *et al.* 2017) and pressure-constrained magnetic equilibria. Fig. 1 shows the profiles and corresponding gradient scale length used in this study. We focus on the time point 6 ms after the ELM crash, where the pedestal is mostly recovered and profiles are almost pre-ELM, with a slightly ($\approx$ 7%) reduced electron temperature at the pedestal top. The dashed lines indicate representative positions for pedestal top/shoulder, an intermediate region, pedestal centre, and pedestal foot, where linear, local instability scans have been performed (see Sec. 4). With pedestal top/shoulder we refer to the radial position ($\rho_{tor}$ = 0.86) just before the increase of temperature and density gradients, where the growth rate spectrum is still clearly distinct from the pedestal centre. Fig. 2 shows the $E \times B$ rotation due to the radial electric field $E_r$ and the corresponding shear. In Fig. 3 the radial profiles of further quantities determining edge physics and microinstabilities are shown: plasma $\beta$ (strongly falls off from pedestal top to foot), collisionality (strongly increases), the gyroradius (decreases), and the safety factor q (increases but exhibits an intermediate flat region).

The microinstabilities that dominate under these physical conditions are characterised with linear, local simulations in the next section.

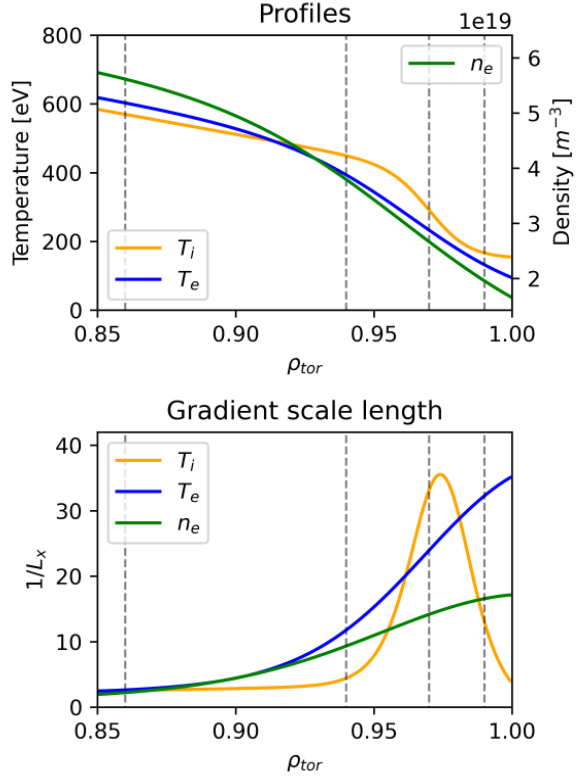

FIGURE 1. Profiles (top) and gradient scale length $1/L_X = -\partial_r X(r)/X$ (bottom) of ion temperature (orange), electron temperature (blue) and density (green) of AUG #31529 6ms after the ELM crash. It is assumed that $n_e = n_i$. The dashed lines indicate positions where instabilities have been characterised in detail (see Sec. 4).

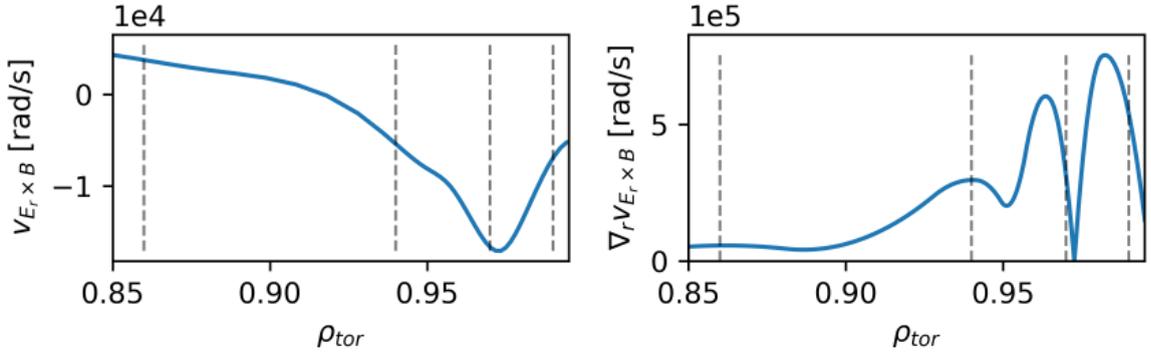

FIGURE 2. $E \times B$ rotation velocity (left) and corresponding shear (right) caused by the edge radial electric field $E_r$.

## 4. Linear instability characterisation

### 4.1. *Local, linear scans*

To characterise the instabilities present in the given pedestal we have performed scans with linear local simulations in the binormal wavenumber $k_y$ from $0.05\rho_i^{-1}$ to $350\rho_i^{-1}$ (corresponding to e.g. $N_{\text{tor}} = [4, 29954]$ at $\rho_{\text{tor}}=0.86$ or $N_{\text{tor}} = [7, 48365]$ at $\rho_{\text{tor}}=0.99$), the radial wavenumber at the outboard midplane $k_{x,\text{center}}$ from -40 to 40 (connected to the ballooning angle $\theta_0 = k_{x,\text{center}}/(\hat{s}k_y)$) and radial position $\rho_{\text{tor}}$ from 0.85 to 1. Fig. 4 shows the growth rate spectra at four positions (pedestal shoulder, intermediate, pedestal centre and pedestal foot) maximised for each $\rho_{\text{tor}}$ and $k_y$ over the ballooning angle. At selected wavenumbers, ballooning angles and radial positions we have additionally performed scans in temperature and density gradients ($\pm 30\%$), collisionality (0 - $18\nu_e^*$) and plasma $\beta$ (0% - 1%). In total about 7000 simulations were performed for the characterisation of linear instabilities.

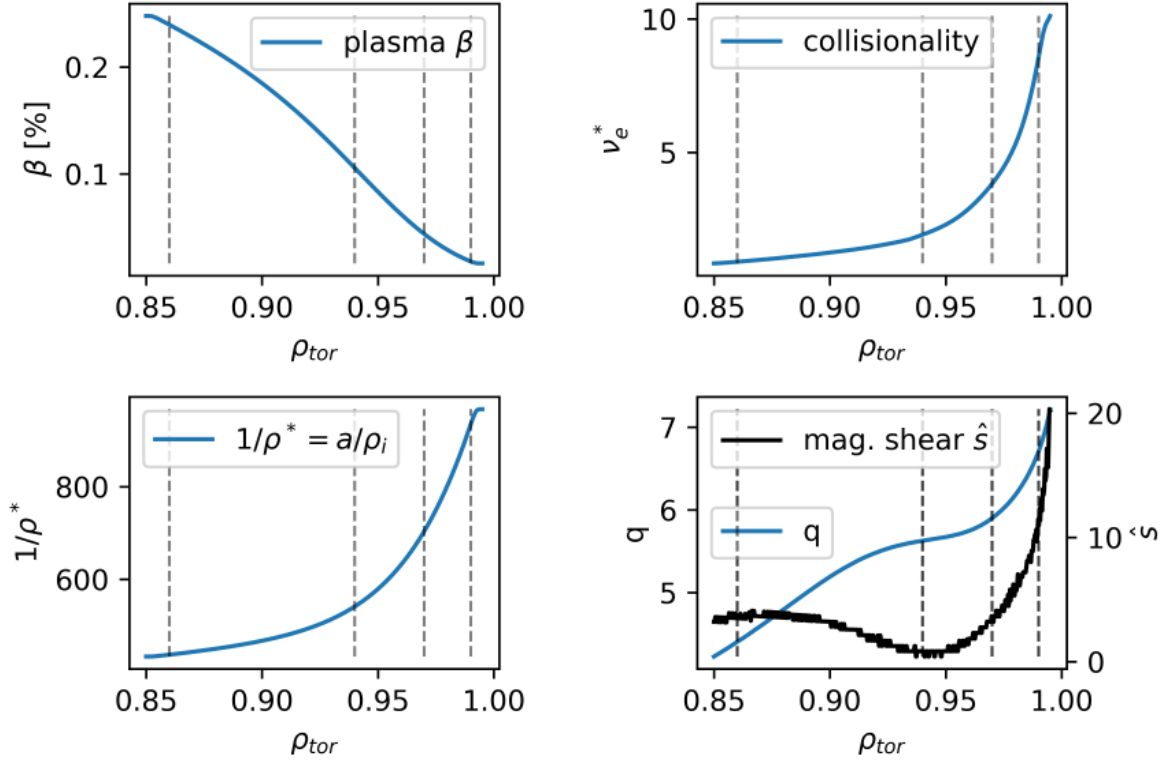

FIGURE 3. Profiles of further relevant quantities influencing microinstabilities and edge turbulence: Plasma $\beta$ (top left), collisionality (top right), $\rho^*$ (bottom left) and safety factor q combined with magnetic shear $\hat{s}$ (bottom right).

We use the following criteria to distinguish between the different instabilities: Parity of the parallel mode structure in ballooning representation (tearing or ballooning), size (ion or electron gyroradius), drift direction (ion or electron diamagnetic), sensitivity on gradients ($T_e$, $T_i$, $n$), dependence on collisionality and plasma $\beta$ as well as diffusivity and heat flux ratios following the fingerprint approach (Kotschenreuther et al. 2019). An electromagnetic mode on scales $> \rho_i$ with tearing parity is called MTM. An electromagnetic mode on scales $> \rho_i$ with ballooning parity and drift in ion diamagnetic direction is called KBM. An electrostatic mode on scales $\approx \rho_i$, which is stabilised by collisionality and is not destabilised by $\nabla T_i$ is called TEM. An electrostatic mode on scales $\approx \rho_i$, which is destabilised by $\nabla T_i$ and propagates in ion diamagnetic direction is called ITG. An electrostatic mode on scales $\lesssim \rho_i$ that is destabilised by $\nabla T_e$ is called ETG. However, it should be emphasised that a clear categorisation of modes on ion scales in the region of very strong drive close to the separatrix is particularly challenging. Different drive mechanisms interact and fuel instabilities with characteristics that do not fall neatly in the mode prototypes developed in the study of core turbulence.

At the pedestal shoulder ($\rho_{\text{tor}} = 0.86$, violet triangles in Fig. 4) we find MTMs on largest scales, TEMs at ion scales, a region of stable wavenumbers and then ETGs on electron scales. At the intermediate position ($\rho_{\text{tor}} = 0.94$, blue squares in Fig. 4), where the magnetic shear has a minimum, a growth rate gap without any clear mode on ion scales exists. On intermediate scales, more ETG modes occur, that were not present at the pedestal shoulder. In the steep gradient region ($\rho_{\text{tor}} = 0.97$, green crosses in Fig. 4) on ion scales we find ITG modes that are close to a KBM transition. At smaller scales, ETG driven modes are present that extend to larger scales towards pedestal centre and foot. At the pedestal foot ($\rho_{\text{tor}} = 0.99$, yellow circles in Fig. 4) on ion scales modes show an ETG/TEM character but tend to be destabilised with increasing collision frequency. The ETG modes at pedestal centre and foot tend to peak increasingly at finite ballooning angle, indicating the presence of toroidal ETG modes (Parisi *et al.* 2020; Told *et al.* 2008). Overall growth rates increase from pedestal top to foot.

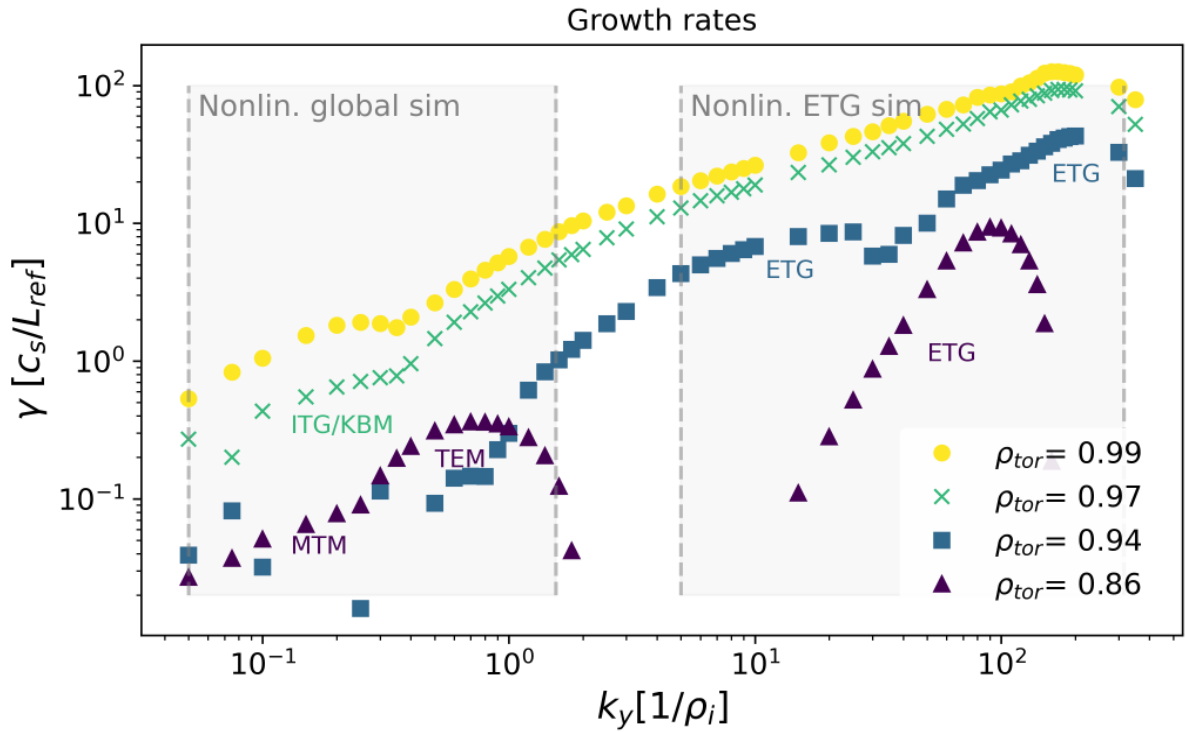

FIGURE 4. Growth rate spectra for four radial positions. Shaded regions indicate the wavenumber ranges used in nonlinear heat flux simulations.

Scans over plasma $\beta$ at different radial positions show that the pedestal in these linear local simulations sits close to a KBM threshold - being closer at the pedestal foot than at the pedestal top (cf. Fig. 5). The closeness to the KBM threshold in the pedestal centre resembles the use of a KBM constraint in the EPED model (Snyder *et al.* 2009) for the prediction of the pedestal width. It has, however, been reported that the radial structure of KBMs may not be compatible with the narrow pedestal region (Hatch *et al.* 2019; Predebon *et al.* 2023), suggesting that the details of the KBM threshold may be skewed by the local approximation.

### 4.2. *Pressure and magnetic shear effect: 2nd stability region*

The low growth rates on ion scales in the intermediate region ($\rho_{\text{tor}} = 0.94$) between pedestal top and steep gradient are caused by the interplay of low magnetic shear and already increased pressure gradient in this region. Their interplay locally stabilises ballooning modes (2nd stability region). This can be shown by a scan over the magnetic shear at this position (see Fig. 6, right plot). At nominal parameters (black crosses) no clear ion scale mode is present, but with increasing magnetic shear $\hat{s}$ (green triangles and yellow circles) the system leaves the 2nd stability region and an ion scale mode becomes unstable. At the pedestal shoulder, where a lot of transport is driven on ion scales, lowering the magnetic shear, unfortunately, does not decrease growth rates significantly, since the pressure gradient is too low to access the 2nd stability region (Fig. 6, left plot green triangles). In contrast, a high magnetic shear lets the pedestal shoulder enter the 1st stability region, where TEMs are suppressed (yellow circles). Further scans with e.g. modified profiles are not within the scope of this paper, but these results illustrate the potential of reducing microturbulence instabilities through careful tailoring of safety factor and pressure profiles.

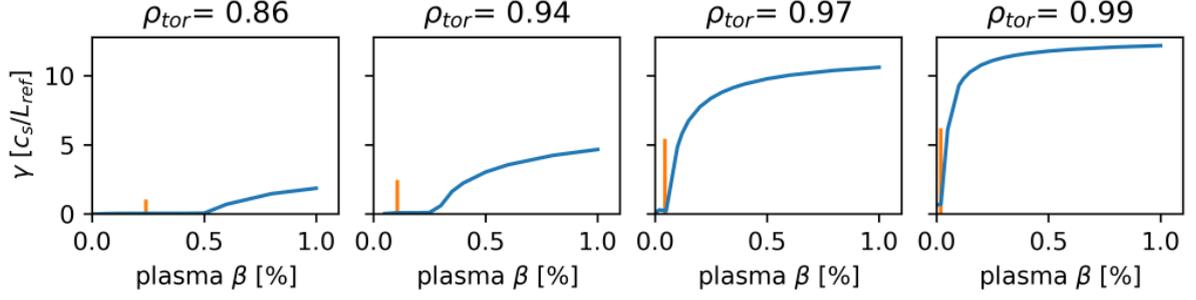

FIGURE 5. Growth rate scan over plasma $\beta$ at $k_y\rho_i = 0.075$ and $k_{x,\text{center}} = 0$ at four radial positions. Nominal $\beta$ value indicated in orange.

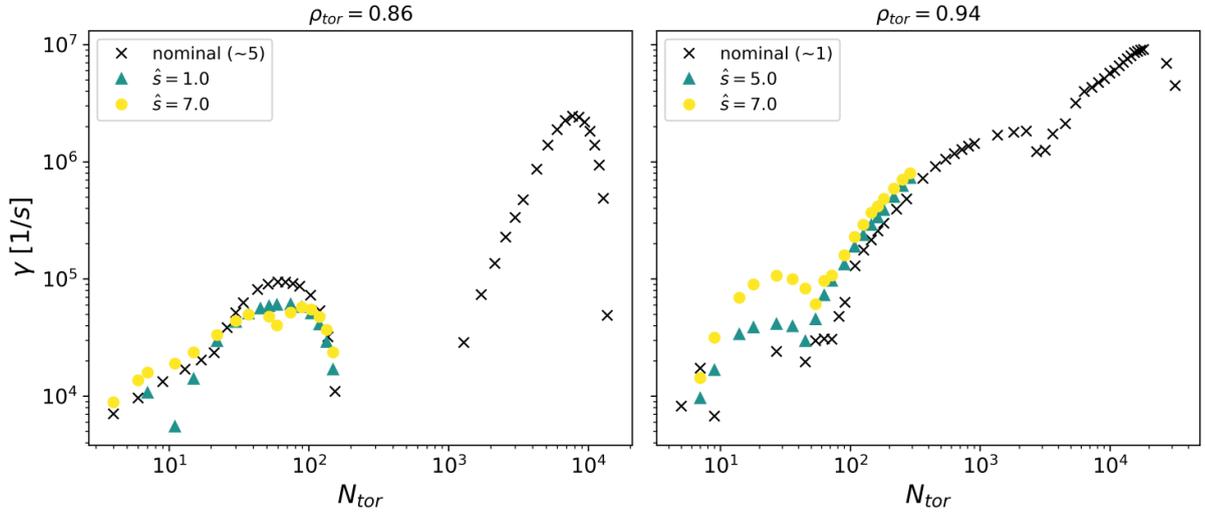

FIGURE 6. Linear, local growth rates as a function of the toroidal mode number $N_{\text{tor}}$ for three magnetic shear ($\hat{s}$) values at two different radial positions (left: pedestal shoulder $\rho_{\text{tor}} = 0.86$, right: low shear region $\rho_{\text{tor}} = 0.94$). Black crosses are nominal growth rates.

    In the next section, we analyse the nonlinear turbulent system that is fuelled by the presented instabilities and compare to what extent linear mode signatures prevail in the nonlinear state. To reduce computational cost ion scales and electron scales are treated separately.

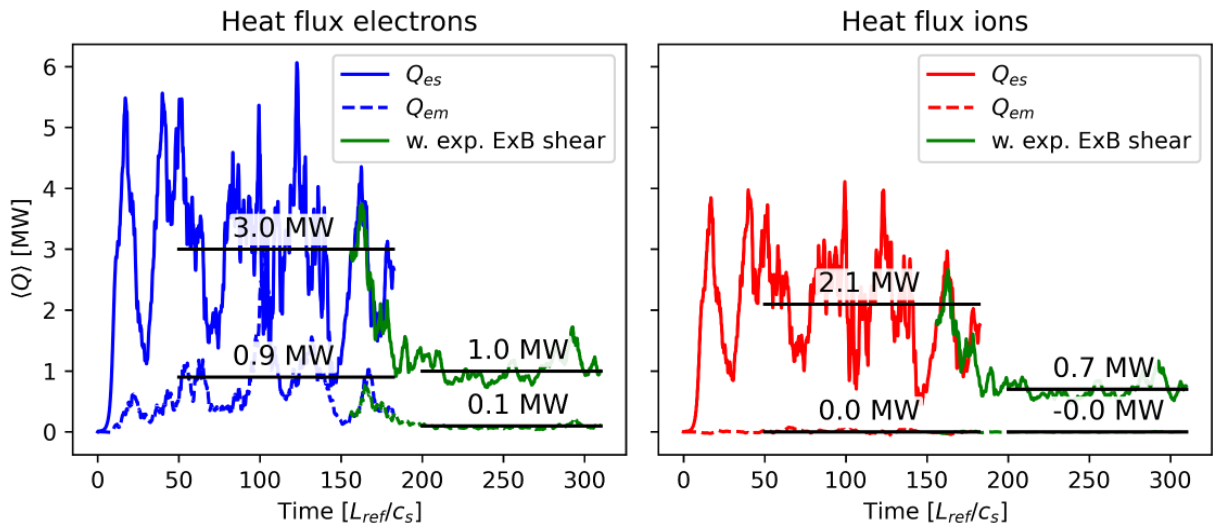

FIGURE 7. Heat fluxes (electrostatic $Q_{es}$ and electromagnetic $Q_{em}$) for electrons (blue, left) and ions (red, right) as a function of time in MW. The green continuations are performed including an external background velocity shear corresponding to experimentally measured $E \times B$ shear.

## 5. Nonlinear heat flux simulations

### 5.1. *Global, ion scale simulations*

To calculate heat fluxes we have performed gradient-driven, global, nonlinear, electromagnetic simulations on ion scales at experimental $\beta$ values, which have been enabled by the code upgrade presented in Sec. 2. These simulations cover $\rho_{\text{tor}} = 0.85 - 0.995$ in radius (almost the full width shown in Fig. 1) and $k_y \rho_i = 0.05 - 1.6$ in binormal wavenumber (corresponding to a toroidal wavenumber $N_{\text{tor}} \approx 4 - 124$), indicated by the left shaded region in Fig. 4. They are two species simulations (Deuterium and electrons) with correct mass ratio ($m_e/m_D = 1/3670$), collisions (Landau collision operator), perpendicular magnetic fluctuations $\bar{A}_{1\parallel}$, but without compressional magnetic perturbations $B_{1,\parallel}$. When indicated, the simulations include the effect of $E \times B$ rotation due to the radial electric field $E_r$ (cf. Fig. 2). For numerical reasons the background flow in GENE is restricted to the toroidal direction. As an approximation to the $E \times B$ shear effect we retain the magnitude of $v_{E \times B}$, but rotate it to be purely toroidal. More details on e.g. grids are specified in the appendix.

The obtained heat fluxes are shown in Fig. 7 as a function of time averaged over real space (flux surface and radius, including the radial buffer zones (see appendix)). In both, ion and electron channel, the electrostatic heat flux dominates. But while the ions show vanishing electromagnetic heat flux, the electron heat flux is about 1/4 electromagnetic. When including $E \times B$ shear in the simulations the heat flux is strongly damped in all channels. The electrostatic by a factor of three and the electromagnetic even stronger.

### 5.2. *Connecting linear and nonlinear results: Frequencies and cross phases*

To identify the dominant turbulent transport mechanisms, results from linear and nonlinear simulations have to be connected, to test if the linearly fastest growing modes remain important in the nonlinear saturated state. This can be achieved by comparing frequencies and cross phases.

Fig. 8 shows the mode frequencies of the linear and nonlinear simulations. The left plot shows the frequency spectrum of the local, linear simulations for three radial positions (pedestal shoulder at $\rho_{\text{tor}} = 0.86$, centre at $\rho_{\text{tor}} = 0.97$ and foot at $\rho_{\text{tor}} = 0.99$), in the co-moving frame of reference. In the pedestal shoulder spectrum (violet triangles) the transition from MTM to TEM is visible and at the pedestal centre (yellow circles) the transition from ITG to TEM/ETG. The right plot shows the frequency spectrum of the global, nonlinear simulation analysed at two positions (pedestal shoulder and centre) overlaid by the local, linear spectra. Usually frequencies from local simulations are specified in a frame co-moving with the $E \times B$ rotation. Since $E \times B$ rotation depends on radial position, frequencies of global simulations are specified in the lab frame and for the comparison of both we transform the local frequencies also to the lab frame. The comparison shows that at the pedestal shoulder and even centre the linear frequencies persist in the nonlinear simulations. This indicates that the linearly fastest growing modes remain important in the saturated turbulent state. However, one important difference appears: At the pedestal shoulder ($\rho_{\text{tor}} = 0.86$) and $k_y \rho_i \approx 0.3$ the nonlinear simulation does not show the mode transition that linear, local simulations present, indicating, that MTMs are suppressed or at least restricted to the very largest scales in global, nonlinear simulations compared to the local, linear ones.

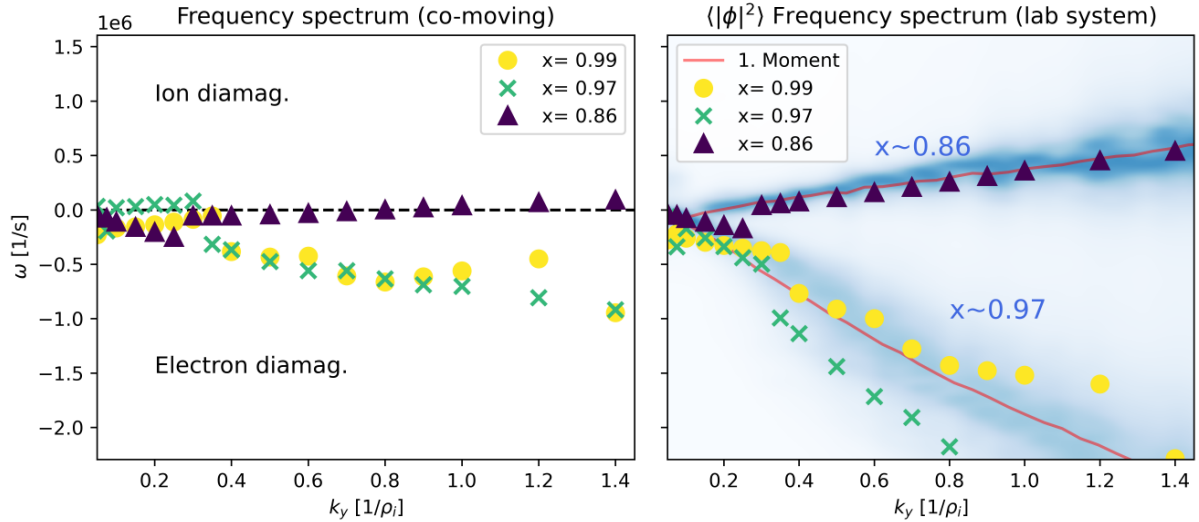

FIGURE 8. Frequency spectrum on ion scales for linear, local simulations in the co-moving frame (left) and global, nonlinear results compared to the linear, local scans in the lab system (right). The dashed line (left) indicates the transition between ion and electron diamagnetic drift direction.

The cross-phases of electric potential and electron density fluctuations (see Fig. 9) support this picture: Linear mode structures survive in the nonlinear simulations, but on largest scales at the pedestal shoulder differences are visible, corroborating the suppression of MTM in global, nonlinear simulations observed in the frequency comparison. Overall, the remarkable agreement seen in the frequency and cross phase comparison between local/linear and global/nonlinear simulations at pedestal shoulder and centre encourages the extension of quasi-linear models to the pedestal region.

### 5.3. *Heat flux profile*

To study how turbulent transport changes across the pedestal we consider the heat flux averaged over time and flux surface as a function of the radius. Fig. 10 shows important aspects of the turbulent heat flux profile in the studied ASDEX Upgrade pedestal. Particular care has to be taken with regard to the normalisation of the heat flux. Due to the strong temperature changes across the pedestal, the commonly used gyro-Bohm heat flux based on mixing-length estimates $Q_{gB} = c_s p_i \rho^{*2} = (T_i/m_i)^{1/2} n_i T_i \rho^{*2}$ changes by two orders of magnitude across the pedestal (top left plot). A modified gyro-Bohm heat flux $Q_{\text{gb,mod}} = Q_{\text{gb}} \times \max{(a/L_{Ti}, a/L_{Te})}^2$, in which the minor radius is replaced by the gradient length as the macroscopic length scale, exhibits strong variations across the pedestal as well. We will therefore focus the analysis on the heat flux in SI units.

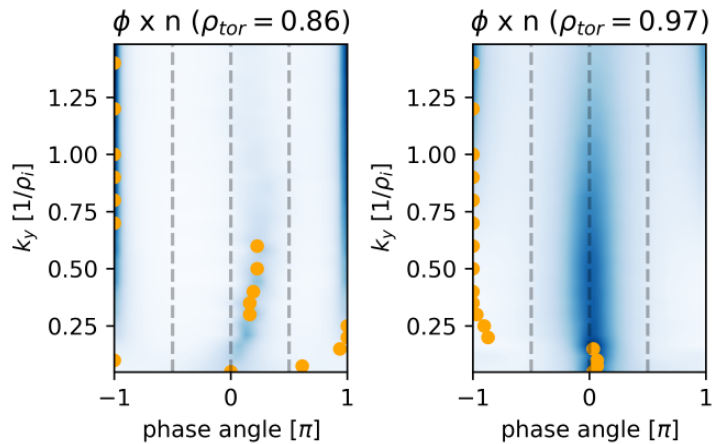

FIGURE 9. Cross-phases of electric potential $\phi$ and electron density fluctuations $n$ from nonlinear simulations (blue background) and linear simulations (orange circles).

Fig. 10 (top right) shows the profiles of the different heat flux components (electrostatic, electromagnetic, electron, ion). The dominant component over most of the pedestal is electrostatic electron heat flux (blue solid line), with the exception of $\rho_{\text{tor}} \approx 0.97$, where the electrostatic ion heat flux has a local peak. This peak corresponds to the ITG/KBM mode identified in the linear analysis, which occurs at the peak of the ion temperature gradient. Interestingly, the turbulent ion scale heat flux is strongly reduced in all channels from pedestal top to pedestal centre and foot - even without $E \times B$ shear. The onset of this reduction coincides with the region of linear stabilisation, discussed in Sec. 4.2.

Fig. 10 (bottom left) illustrates the influence of $E \times B$ shear on the heat flux profiles: It reduces the ion scale heat flux strongly and widens the region of almost vanishing turbulent heat flux.

Fig. 10 (bottom right) shows that the turbulent heat flux on electron scales behaves oppositely across the pedestal compared to the ion scale turbulent heat flux: It vanishes at the pedestal top and strongly increases down the pedestal (stars).

A comparison to power balance and neoclassical calculations shows qualitative agreement in heat flux structure and trends (bottom right plot). The electron channel is dominant throughout the pedestal and roughly constant. Our gyrokinetic simulations reveal that while the total electron heat flux remains constant, it transitions in scale. At the pedestal top/shoulder it is driven by ion scale TEM/MTM turbulence while at the pedestal centre and foot it is driven by small-scale ETG turbulence. The region around $\rho_{\text{tor}} = 0.92$ where the total gyrokinetic electron heat flux reaches a minimum likely indicates a limitation of our separate scale ansatz. At this location heat flux is likely driven by scales that are neither resolved in our global simulations nor in the electron scale simulations (cf. Fig. 4) and would require multi-scale simulations to be properly resolved. The ion channel contributes substantially to the total heat flux at the pedestal top/shoulder and reduces to neoclassic values towards the pedestal centre. At the pedestal foot our gyrokinetic simulations do not show the increase of ion heat flux suggested by power balance. The differences in the grey-shaded region are possibly due to increased measurement uncertainties in the ion profile that affect both power balance (see (Viezzer *et al.* 2020)) and gyrokinetic simulations.

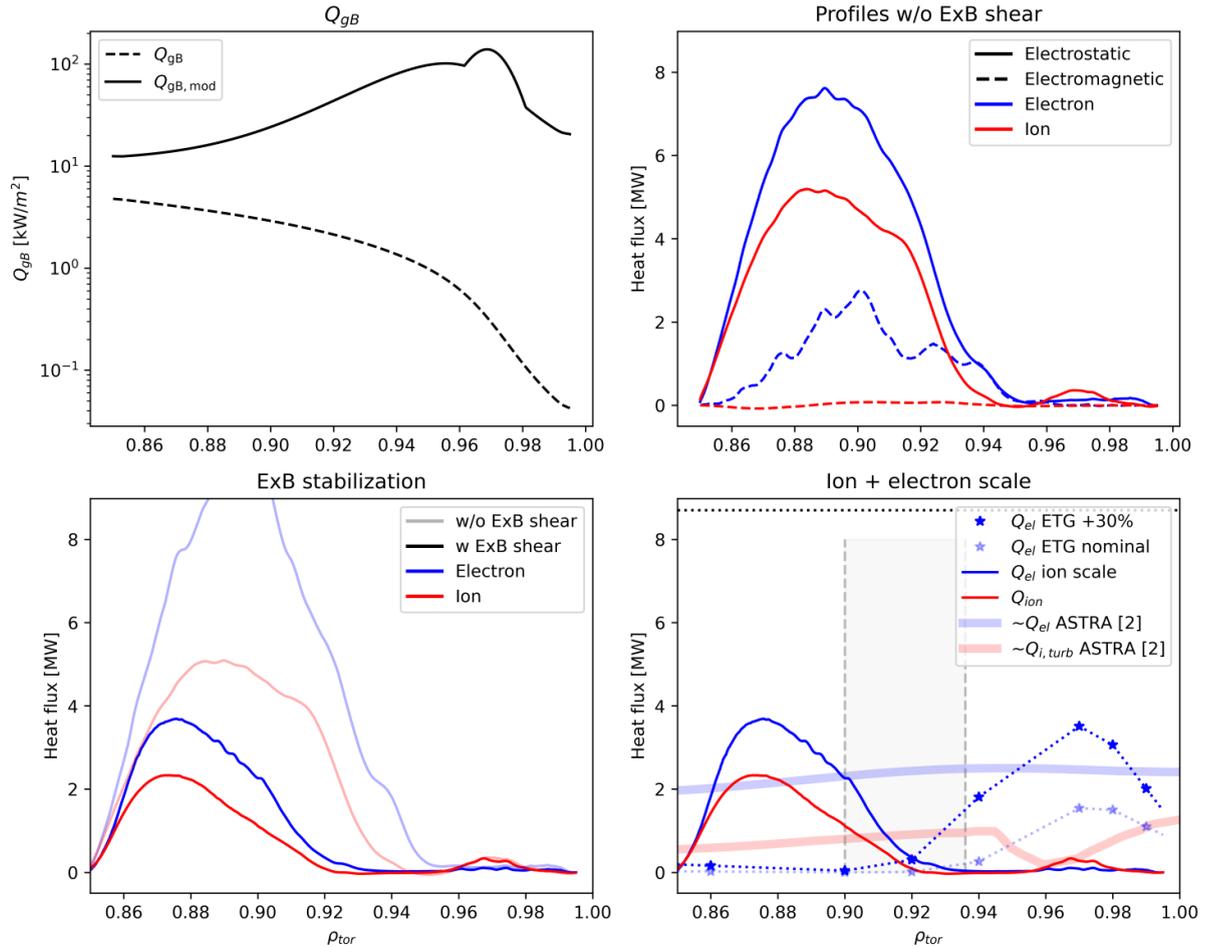

FIGURE 10. Turbulent heat flux profile in an ASDEX Upgrade pedestal from pedestal top to foot. Top left: Gyro-Bohm heat flux profile. Top right: Components of the ion scale heat flux profile without $E \times B$ shear. Bottom left: Total heat flux ($Q_{es}+Q_{em}$) due to ion scale turbulence with and without $E \times B$ shear. Bottom right: Total heat flux due to ion scale turbulence from global simulations (red and blue solid line) as well as ETG heat fluxes from local simulations at nominal values (light blue stars) and increased electron temperature gradient (dark blue stars) compared with power balance calculations (broad lines). Region of increased measurement uncertainty in grey.

For the power balance comparison, the turbulent ion heat flux component was estimated by subtracting neoclassic heat flux calculated with NEOART from the ASTRA result (both from (Viezzer *et al.* 2020)). The total ion heat flux due to power balance is constant (Viezzer *et al.* 2020) and the minimum in the turbulent heat flux is compensated by an increased neoclassical heat flux, which has a roughly constant diffusivity across the pedestal so that its heat flux follows the increasing ion temperature gradient.

The following picture emerges for the investigated scenario: Turbulent heat flux at the pedestal top is dominated by electrostatic TEMs with electromagnetic MTM contributions. The interplay of low magnetic shear and increasing pressure gradient stabilises modes locally before the steep gradient region ($\rho_{tor} \approx 0.94$) and reduces heat flux. $E \times B$ shear suppresses heat flux strongly in all channels and widens the region of vanishing ion scale heat flux. Turbulent electron heat flux changes scale across the pedestal and turbulent ion heat flux strongly reduces towards the pedestal centre.

## 6. Discussion & Conclusions

We have presented a gyrokinetic analysis of an ASDEX Upgrade ELMy H-mode pedestal. The most unstable microinstabilities from just inside the pedestal top to foot were characterised with extensive linear, local scans. At the pedestal top/shoulder MTM, TEM, and ETG were found. In the intermediate region before the pedestal centre modes on ion scales are stabilised by the interplay of magnetic shear and pressure gradient, while on intermediate electron scales additional ETG modes become unstable. In the pedestal centre ITG modes close to the threshold to KBMs were observed and at the pedestal foot modes with TEM/ETG character that are destabilised with increasing collision frequency are present. With nonlinear, global, electromagnetic ion scale simulations and nonlinear, local ETG simulations we have analysed the heat flux in the pedestal - resolved in radius and scale. The ion scale simulations are enabled by an upgrade of the GENE code (cf. Sec. 2). We find TEM-driven turbulence with electromagnetic components due to MTMs to be dominant at the pedestal top/shoulder. A combination of linear stabilisation and $E \times B$ shear suppresses ion scale turbulence towards the steep gradient region. While the turbulent electron heat flux is picked up by small-scale ETG modes, the ion channel reduces to neoclassic heat flux levels.

The global electromagnetic simulations presented in this study are among the most realistic pedestal turbulence simulations performed to date. Together with dedicated local ETG simulations and the extensive linear instability characterisation, they help to confirm the important role of $E \times B$ shear stabilisation for pedestal turbulence and demonstrate the transition in scale of electron turbulence from ion scales at pedestal top to electron scales in the steep gradient region. This well-resolved characterisation of turbulence across a pedestal supports the development of reduced models for edge turbulence and the understanding of the L-H transition.

Two limitations of our current approach should be addressed in future work: The separation of ion and electron scale simulations prohibits any mutual scale interaction and excludes a range of wavenumbers in our setup. Multi-scale simulations would promise new insights into how turbulent transport transforms across the pedestal. Furthermore, the use of field-aligned coordinates strictly restricts our simulation domain to the region of closed flux surfaces. Approaches that can cross the separatrix would be well suited to expand the study of pedestal foot turbulence.

## Acknowledgements

The authors thank Eleonora Viezzer, Felix Wilms, Sergei Makarov, Teobaldo Luda di Cortemiglia and Ben Chapman-Oplopoiou for fruitful discussions and suggestions. Simulations for this work have been performed on the HPC systems Cobra and Raven at the Max Planck Computing and Data Facility (MPCDF) as well as Marconi at CINECA. This work has been carried out within the framework of the EUROfusion Consortium, funded by the European Union via the Euratom Research and Training Programme (Grant Agreement No 101052200 - EUROfusion). Views and opinions expressed are however those of the author(s) only and do not necessarily reflect those of the European Union or the European Commission. Neither the European Union nor the European Commission can be held responsible for them.

## Declaration of interest

Competing interests: The authors declare none.

**Appendix**

6.1. *Simulation parameters*

6.1.1. *Linear, local simulations*

- 2 species, experimental $\beta$, realistic electron to ion mass ratio $m_e/m_D = 1/3670$, Landau collision operator. $E \times B$ shear was not used to avoid Floquet modes.
- Resolution: $n_x = 18$, $n_{ky} = 1$, $n_z = 36$, $n_v = 32$, $n_w = 16$.
- Box size: lv=3.1, lw=11.
- Convergence tests with increased parallel resolution ($n_z = 144$) and increased velocity space resolution ($n_v = 128$, $n_w = 32$) were performed.

6.1.2. *Nonlinear, local ETG simulations*

- 1 kinetic species (electrons), adiabatic ions, experimental $\beta$, Landau collision operator, no $E \times B$ shear.
- Resolution: $n_x = 512$, $n_{ky} = 64$, $n_z = 72$, $n_v = 32$, $n_w = 16$.
- Box size: lv=3.45, lw=14.23, lx=115.
- Tests with $E \times B$ shear were performed, finding a heat flux reduction of $< 15\%$ by $E \times B$ shear.

6.1.3. *Nonlinear, global, ion scale simulations*

- 2 species, experimental $\beta$, realistic electron to ion mass ratio $m_e/m_D = 1/3670$, Landau collision operator. With $E \times B$ shear when indicated.
- Resolution: $n_x = 512$, $n_{ky} = 32$, $n_z = 48$, $n_v = 32$, $n_w = 16$.
- Box size: lv=3.45, lw=14.23, lx=72.
- Boundary conditions: Dirichlet with radial buffer zones (5% percent of domain at both boundaries), in which the distribution function is damped by fourth-order Krook operators.
- Performed with block-structured velocity grids (Jarema *et al.* 2017) with 4 blocks.
- Performed in single-precision floating-point format.